\documentclass[12pt,preprint]{aastex}

\def\gsim{\,\lower3pt\hbox{$\sim$}\llap{\raise2pt\hbox{$>$}}\,}
\def\lsim{\,\lower3pt\hbox{$\sim$}\llap{\raise2pt\hbox{$<$}}\,}
\newcommand{\be}{\begin{equation}}
\newcommand{\ee}{\end{equation}}
\newcommand{\bex}{\begin{equation}\notag}
\newcommand{\eex}{\end{equation}\notag}
\newcommand{\bea}{\begin{eqnarray}}
\newcommand{\eea}{\end{eqnarray}}
\newcommand{\beax}{\begin{eqnarray*}}
\newcommand{\eeax}{\end{eqnarray*}}
\newcommand{\ba}{\begin{array}}
\newcommand{\ea}{\end{array}}

\begin{document}

\title{Radiation Hydrodynamics in Solar Flares}
\shorttitle{Hydrodynamics in Flares}


\author{G.~H. Fisher}
\affil{Institute of Geophysics and Planetary Physics, Mail Code L-413,
Lawrence Livermore National Laboratory, Livermore, CA 94550}


\begin{abstract}
Solar flares are currently understood as the explosive release of energy 
stored in the form of stressed magnetic fields.  In many cases, the released 
energy seems to take the form of large numbers of electrons accelerated to 
high energies (the nonthermal electron ``thick target'' model), 
or alternatively 
plasma heated to very high temperatures behind a rapidly moving conduction 
front (the ``thermal'' model).  The transport of this energy into the remaining 
portion of the atmosphere results in violent mass motion and strong emission 
across the electromagnetic spectrum.  Radiation processes play a crucial role 
in determining the ensuing plasma motion.

One important phenomenon observed during flares is the appearance in 
coronal magnetic loops of large amounts of upflowing, soft X-ray emitting 
plasma at temperatures of $1-2\ \times 10^7$ [K].  It 
is believed that this is due to 
chromospheric evaporation, the process of heating cool ($T \sim 10^4$ [K]) 
chromospheric material beyond its ability to radiate.  Detailed calculations 
of thick target heating show that if nonthermal electrons heat the 
chromosphere directly, then the evaporation process can result in explosive 
upward motion of X-ray emitting plasma if the heating rate exceeds a threshold 
value.  In such a case, upflow velocities approach an upper limit of roughly 
$2.35 c_s$ as the heating rate is increased beyond the threshold, 
where $c_s$ is the 
sound speed in the evaporated plasma.  This is known as explosive
evaporation.  If the flare heating rate is less than the threshold, 
evaporation takes place indirectly through thermal conduction of heat 
deposited in the corona by the energetic electrons.  Upflows in this case are 
roughly 10 to 20\% of the upper limit.  Evaporation by thermal model heating 
always takes place through thermal conduction, and the computed upflow speeds 
seem to be about 10\% to 20\% of the upper limit, independent of the energy 
flux.

The pressure increase in the evaporated plasma for either the thick 
target or thermal model leads to a number of interesting phenomena in the 
flare chromosphere.  The sudden pressure increase initiates a downward moving 
``chromospheric condensation'', an overdense region which gradually decelerates 
as it accretes material and propagates into the gravitationally stratified 
chromosphere.  Solutions to an equation of motion for this condensation shows 
that its motion decays after about one minute of propagation into the 
chromosphere.  When the front of this downflowing region is supersonic
relative to the atmosphere ahead of it, a radiating shock will form.  If the 
downflow is rapid enough, the shock strength should be sufficient to excite UV 
radiation normally associated with the transition region, and furthermore, the 
radiating shock will be brighter than the transition region.  These results 
lead to a number of observationally testable relationships between the optical 
and ultraviolet spectra from the condensation and radiating shock.
\end{abstract}

\section*{I.  Introduction}
\label{section:intro}

Solar flares are rather violent and extremely complicated phenomena, and it 
should be made clear at the outset that a physically complete picture 
describing all aspects of flares does not exist.  From the wealth of data 
which is available, it is apparent that many different types of physical 
processes are involved during flares: energetic particle acceleration, rapid 
magnetohydrodynamic motion of complex field structures, magnetic reconnection, 
violent mass motion along magnetic field lines, and the heating of plasma to 
tens of millions of degrees, to name a few.  The goal of this paper will be to 
explore just one aspect of solar flares, namely the interaction of 
hydrodynamics and radiation processes in fluid being rapidly heated along 
closed magnetic field lines.  The models discussed are therefore necessarily 
restrictive, and will address only a few of the observed or observable 
phenomena.

In Section II, the structure of the solar corona is discussed, and 
evidence arguing for the use of the one dimensional loop model is presented.  
In Section III, two popular models for energy release in flares are presented, 
along with some brief discussion of pertinent observations.  Section IV 
describes the physics included in the numerical simulations of radiative 
hydrodynamic response of flare loops to energy release.  Specific results of 
such simulations and related analytical calculations are given in Sections V, 
VI, and VII, on the phenomena of chromospheric evaporation, the formation and 
propagation of very dense chromospheric condensations, and emission from 
radiating shocks, respectively.  Finally, new results concerning possibly 
observable features of chromospheric condensations and radiating shocks are 
summarized in the final Section, VIII.

\section*{II.  Solar Atmospheric Structure:  The Loop Model}
\label{section:loopmodel}

One of the most important discoveries to come out of space based observations 
of the Sun's corona is the degree to which the structure of the upper solar 
atmosphere is determined by magnetic fields.  In the absence of magnetic 
fields, a balance between gravity and gas pressure would result in a 
spherically symmetric distribution of matter.  However, spatially resolved 
observations of the Sun's coronal X-ray emission show this is far from the 
case [1].  There are large patches of the Sun's surface which appear to show 
essentially no X-ray emission.  These are called coronal holes, which are 
believed to be regions where the magnetic field lines from the surface open 
into interplanetary space.  Gas pressure gradients along these diverging field 
lines are believed to drive the solar wind against the opposing gravitational 
force.  As a result of its rapid expansion, the coronal material has a low 
density, and it therefore emits few X-rays.  Most of the solar surface, 
however, appears to covered by closed magnetic field lines:  A single magnetic 
field line connects two different points of the Sun's surface, denoted as
magnetic footpoints.  Coronal X-ray emission from the magnetically closed 
region seems to consist of emission from individual flux tubes, or bundles of 
adjacent field lines, each with its own characteristic brightness.  The 
brightest and most visible flux tubes occur in active regions, where the 
photospheric magnetic field tends to be both strong and complex.  The 
difference in observed brightness between individual flux tubes can be orders 
of magnitude, implying that a large range in coronal density also exists 
between individual coronal flux tubes, or coronal loops as they are often 
called.  This leads one to conclude that for the purposes of describing the X-
ray emitting corona, the coronal loop can be considered as a basic unit of 
structure.  Furthermore, one can take advantage of the wide range in inferred 
loop densities to note that the magnetic forces must be strongly dominant over 
those from gas pressure gradients:  two different coronal loops of the same 
height would have roughly the same gas pressure if magnetic forces were weak 
or comparable to pressure gradients and gravity.  Thus pressure gradients tend 
to drive mass motion only along the direction of the field lines.

In addition to being the most important agent in determining the 
structure of the upper solar atmosphere, magnetic fields are also believed to 
be the energy source for many dynamic solar phenomena, including, but not 
restricted to, solar flares.  It is generally believed that flares occur when 
stresses in the magnetic fields which accumulate through motion of the 
magnetic footpoints (by differential rotation of the Sun's surface, for 
example) result in rapid and explosive reconfiguration of the field lines into 
a lower energy state.  The energy that is released from this explosive 
phenomenon takes a wide variety of forms, including the escape into space of 
energetic particles, violent mass motion along coronal flux tubes, and strong 
emission across the electromagnetic spectrum from radio waves to $\gamma$-rays.

The morphology of solar flares is in general very complex, and it is 
often difficult to make meaningful generalizations.  Nevertheless, solar 
flares are often divided into two types depending on their geometric structure 
[2].  The compact flare apparently takes place in one or a few small closed 
coronal loops which do not seem to alter their size or shape during the flare, 
but which do become very bright in soft X-rays as the flare progresses.  This 
implies that over the course of the flare, the magnetic pressure always 
dominates over the rapidly increasing gas pressure, as the loop would 
otherwise be blown apart as it filled up with hot plasma.  Another type of 
flare commonly seen is called the two ribbon flare, which seems to consist of 
a wide arcade of bright coronal loops, the chromospheric footpoints of which 
form two bands or ribbons when viewed in visible emission such as H$\alpha$.  
The coronal loop arcade and the H$\alpha$ ribbons combine 
to form a structure much like a 
covered wagon.  The H$\alpha$ ribbons are often observed to 
brighten and move apart 
over the course of the flare. [2]  Two ribbon flares also seem to be 
associated with some open field lines, [3] and some models propose that the 
expansion of the H$\alpha$ ribbons can be understood in terms of 
reconnection of open field lines into closed loop structures. [4]

Because of its apparently much simpler nature, most efforts at modeling 
the radiative hydrodynamic response to explosive flare energy release have 
focused on the compact flare picture:  Flare energy is released within a 
single closed coronal loop model, whose size and shape are not allowed to 
change over the course of a calculation.  Mass motion is allowed to occur only 
along the direction of the magnetic field, owing to the inability of the
transverse pressure gradient to push the field lines out.  In the following 
section, aspects of observations which have bearing on models for flare 
heating and energy transport are reviewed, along with two different pictures 
for how the flare energy is released.

\section*{III.  Flare Observations and Energy Release Models}
\label{section:flareobs}

Studies of radiated emission indicate that flares occur in two different 
stages.  First, during the relatively short impulsive phase, emission seen in 
hard X-rays (meaning photon energies greater than 20 [KeV]) consists of many 
individual spikes ranging in duration from tens of milliseconds to tens of 
seconds.  Impulsive phase emission is also seen in UV lines such as OV [5], 
optical line emission such as H$\alpha$ [6], and frequency integrated 
EUV emission in the 10 to 1030 [\AA] range [7].  
The impulsive phase is followed by the longer 
gradual or thermal phase, which is characterized by bright X-ray emission from 
an optically thin thermal plasma with temperatures of roughly 
$1-2\ \times 10^7$ [K], 
gradually decaying over time scales of 20 minutes to one hour.  It has been 
recently shown by observations from the Solar Maximum Mission (SMM) that blue 
shifts in CaXIX X-ray line emission indicating 
significant upflows are present 
during the impulsive phase.  Upflows seem to begin and end simultaneously with 
hard X-rays.  During this same period of time, the thermal soft X-ray emission 
measure is rapidly increasing, and reaches its peak at about the same time as 
cessation of hard X-ray emission. [8]  The temporal relationship between 
impulsive phase emission, blue shifts, and the soft X-ray emission measure 
suggests that there is some causal connection between them.  In the context of 
the simple compact loop flare, two models of how the flare energy is released 
into the atmosphere have been proposed.

The thick target model [9,10] assumes that all of the impulsive phase 
flare energy is released (in an unspecified manner) into the acceleration or 
energization of energetic nonthermal electrons, which then bombard the 
atmosphere along the coronal loop.  Collisions of the energetic electrons with 
ions in the atmosphere then produce the observed hard X-rays by nonthermal 
bremstrahlung, and collisions with the ambient electrons produce heating.  The 
upper chromosphere, in this scenario, is heated sufficiently to become 
thermally unstable, and then heats up rapidly to coronal temperatures.  This 
``evaporated'' material then expands upward into the coronal loop, giving rise 
both to the observed blue shifted emission seen in CaXIX, as 
well as the rapid 
rise in the total soft X-ray emission measure.  In the meantime, the electrons 
penetrating further down into the chromosphere heat it as well, but below the 
threshold needed for evaporation.  This portion of the chromosphere quickly 
reaches quasi-steady equilibrium between particle heating and radiative 
losses.  The resulting increase in radiative losses is then alleged to account 
for the impulsive component of the optical, UV, and EUV emission.

The thermal model, on the other hand, assumes that the energy release 
takes place by heating some coronal plasma to very high temperatures 
(approximately $10^8$ [K]), and that this energy is then transported to 
the rest 
of the flare loop through the propagation of a thermal conduction front.  
During the initial stage, before the conduction front has had a chance to 
reach the chromosphere, the hot thermal plasma behind the front is alleged to 
account for the impulsive phase hard X-rays through the emission of thermal 
bremstrahlung.  When the conduction front reaches the chromosphere, it begins 
the process of evaporation, and once again the evaporating material is alleged
to account for both the observed blue shifts of CaXIX and the increase 
in the 
soft X-ray emission measure.  The emission of UV, optical, and EUV radiation 
in the thermal model occurs either through the escape of a small fraction of 
the most energetic electrons through the conduction front into the 
chromosphere, or through heating of the chromosphere when the front impacts it.

\begin{figure}
\includegraphics[width=6.5in]{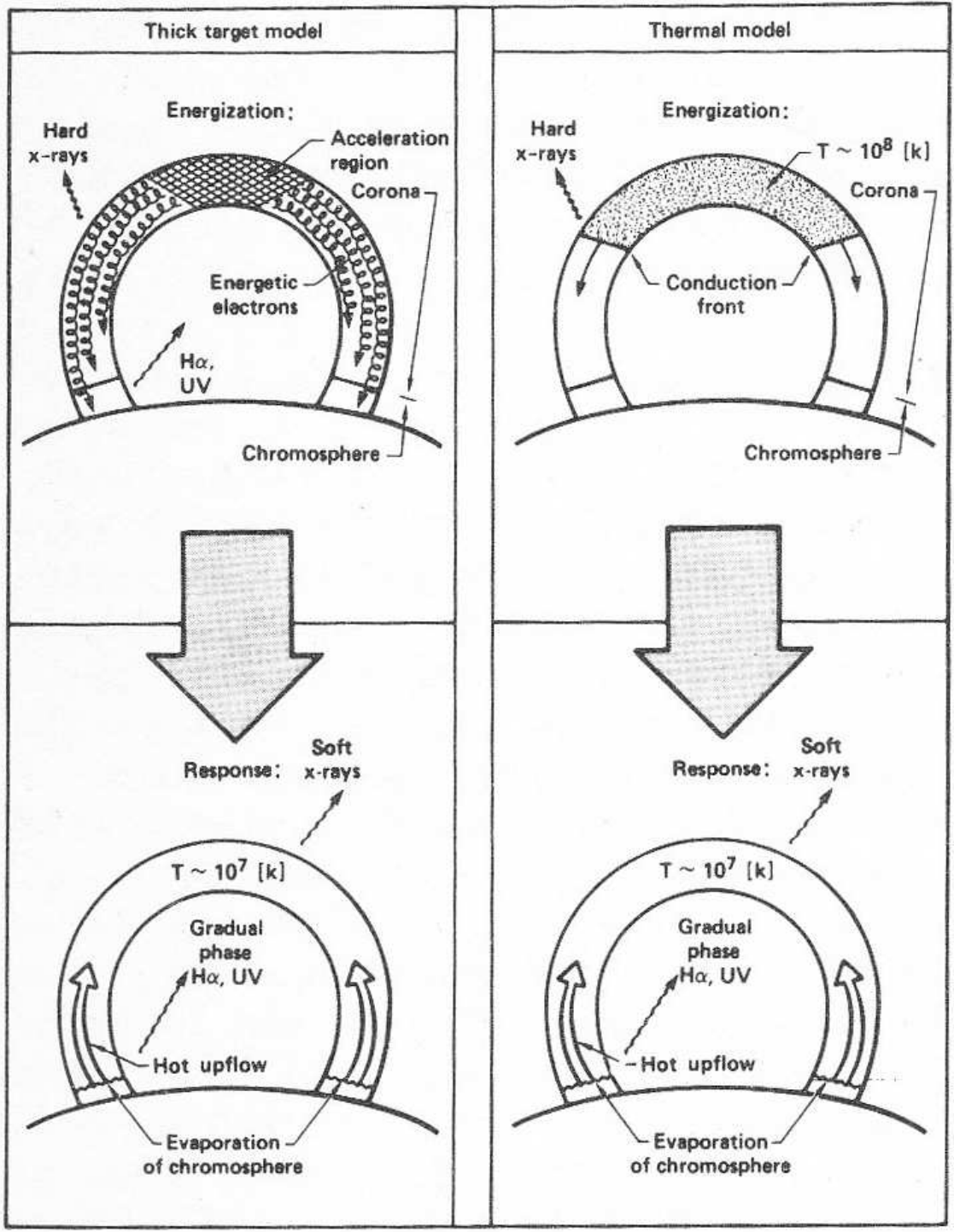}
\caption{
Schematic diagram 
showing how loop is energized in 
the thick target and thermal 
model scenarios.  Note response 
is alleged to be same for both 
models.
}
\label{fig:figure1}
\end{figure}

Both the thick target and thermal models, which are illustrated 
schematically in Fig. 1, represent extreme  and idealized assumptions about 
how the impulsive phase flare energy is released.  In the thick target case, 
the flare energy is released over a wide range of column depth, while the 
energy release in the thermal model is localized to the narrow column depth 
region of the preflare corona.  Nevertheless, these two models are useful 
conceptually, as certain aspects of the observations are easily understood in 
terms of them.  In addition, the thick target and thermal models produce 
heating functions which can easily be computed when doing radiation-
hydrodynamic modeling of the response of a compact loop structure to impulsive 
phase heating.  The goal of doing such modeling is to investigate in detail 
just exactly what the response of a model loop is to each form of flare 
heating, to discover which aspects of the models agree quantitatively with 
observations, and to investigate differences in the computed response which 
might suggest observational tests for discriminating between the two heating 
models.

\section*{IV.  Numerical Modeling of Flaring Loops}

\noindent
\underline{A.  The Equations}

It is clear that the hydrodynamic and radiative response of a model loop 
atmosphere to flare heating is a complicated, non-linear problem.  Therefore, 
much effort has been devoted to numerical solutions of both the hydrodynamic 
response of the flare loop [11-20] as well as calculations of the radiative 
output of semi-empirical flare models [21,22].  It is clear, however, that the 
most complete approach is the self-consistent solution of both the equations 
of hydrodynamics and radiation transport in the flaring loop.  The first 
successful treatment of radiation-hydrodynamics in flare loops was developed 
by MCCLYMONT and CANFIELD [23].  These methods have since been used to study 
nonlocal radiation transport effects in quiet loop models (CANFIELD, FISHER, 
and MCCLYMONT [24]), the linear stability of quiet loop models (MCCLYMONT and 
CANFIELD [25]), the nonlinear evolution of unstable loop models (AN, 
CANFIELD, FISHER, and MCCLYMONT [26]), and the response of loops to both 
thick target (FISHER, CANFIELD, and MCCLYMONT [27,28,29]) and thermal (FISHER 
[30]) models of flare heating.  The details of the physics included in the 
modeling of flaring loops, as well as some brief discussion of the numerical 
techniques, will be described in this section.

The compact loop is assumed to be semi-toroidal, with footpoints at 
either end imbedded in the photosphere.  The loop is also assumed to be 
symmetric about the apex, so that only half the loop is actually modeled.  The 
initial loop structure contains a corona, transition region, chromosphere, and 
the upper portion of the photosphere, all in hydrostatic and energetic 
equilibrium, as described by [26].  The cross sectional area of the loop is 
assumed not to vary along its length.  The model loop geometry is shown 
schematically in Fig. 2.  The equation of motion for the fluid within the loop 
is
\be
m \partial v / \partial t = - \partial P / \partial N + m g_{\parallel} 
+ \partial / \partial N [ (4/3) n \eta \partial v / \partial N ]
\label{equation:equation1}
\ee
where the independent spatial variable $N$ (the column depth) is given by
\be
N = \int_0^z\ n dz'\ ,
\label{equation:equation2}
\ee
measured from the loop apex.  The quantity n is the density of equivalent 
hydrogen atoms ($i.e.$ $n = n_H + n_p$).  The quantity m is the mean mass per 
hydrogen nucleus in the solar atmosphere.

The continuity equation is written, for computational purposes as the pair
of equations
\be
n = (\partial z / \partial N)^{-1}\ ;\ \partial z / \partial t = v\ ,
\label{equation:equation3}
\ee
from which the conventional form is easily obtained.

The energy equation is written
\be
\partial \epsilon / \partial t = Q_a + Q_{fl} - R + C - 
P \partial v / \partial N + (4/3) n \eta | \partial v / \partial N |^2\ ,
\label{equation:equation4}
\ee
where the quantity $\epsilon$, the internal energy per equivalent H-atom, is
given by
\be
\epsilon = 3/2 (1+y+x) kT + \sum_{\ell} \epsilon_{\ell} \phi_{\ell}
\label{equation:equation5}
\ee
where $\epsilon_{\ell}$ is the energy of a hydrogen atom in quantum state 
$\ell$, measured from its ground state, $\phi_{\ell}$ is the fractional
population of state $\ell$ ($\phi_{\ell} = n_{\ell} / n$), $y$ is the number
of non-hydrogenic atoms per H-atom, and $x$, the ionized fraction, is the
number of free electrons per H-atom.  The quantity $Q_a$ here is the quiescent
energy needed to keep the preflare loop in energetic equilibrium, $Q_{fl}$ is
the flare heating function, $R$ is the contribution to the radiative losses
(or radiative heating), and $C$ is the contribution to heating or cooling
from thermal conduction.  These quantities will be described in more detail
in the next subsection.  The fifth term in (\ref{equation:equation4}) is just
the compression term, and the last term in both (\ref{equation:equation1})
and (\ref{equation:equation4}) includes the effects of viscosity; the
viscosity $\eta$ contains both a physical component and pseudoviscosity
component used for handling shocks.

\begin{figure}
\includegraphics[width=6.5in]{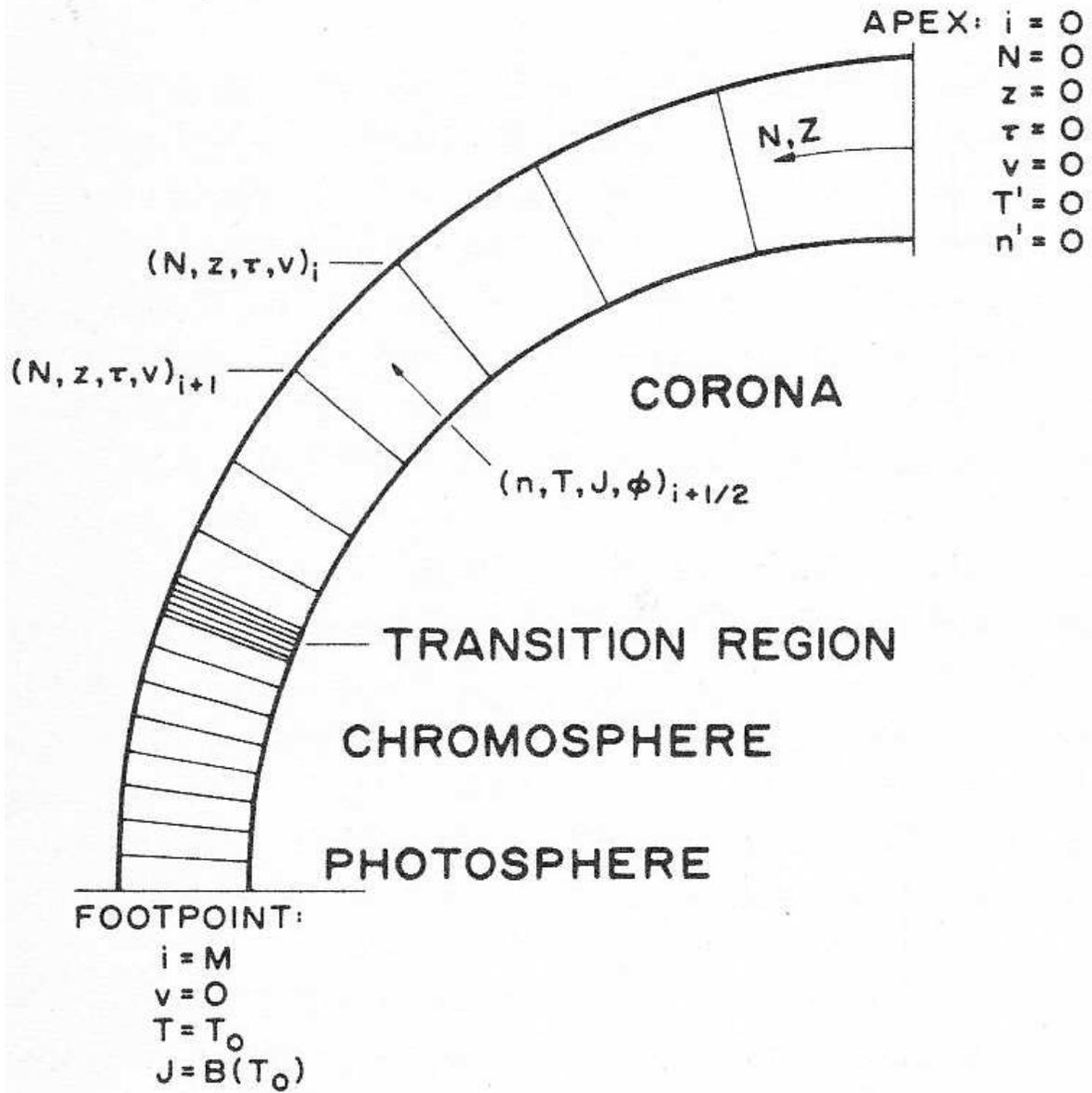}
\caption{
Schematic diagram
showing numerical model of
symmetric flare loop.  This drawing originally appeared in McClymont and
Canfield (1983; ApJ 265, 483)
}
\label{fig:figure2}
\end{figure}

In addition to the hydrodynamic equations, the radiation transport and 
atomic rate equations for a two-level-plus-continuum hydrogen atom are also 
solved.  This is necessary in order to self-consistently determine the ionized 
fraction $x$ and the optically thick hydrogenic losses in the chromosphere.  
The radiation transport equation is reduced to the probabilistic form [31],
\be
d/d \tau_{u \ell} ( J_{u \ell} - S_{u \ell} ) = - 2 p_{u \ell} ^{1/2} 
d / d \tau_{u \ell} (p_{u \ell} ^{1/2} S_{u \ell} )
\label{equation:equation6}
\ee
where the optical depth $\tau_{u \ell}$ is related to the atomic 
photoabsorption cross-section $\sigma_{u \ell}$ by
\be
d \tau_{u \ell} / d N = \sigma_{u \ell}\ .
\label{equation:equation7}
\ee
The quantities $J_{u \ell}$ and $S_{u \ell}$ represent the mean intensity
and source function in transition $u \ell$, respectively, and $p_{u \ell}$
is the single flight escape probability for a photon at optical depth
$\tau_{u \ell}$.  The atomic rate equations are
\be
\partial \phi_i / \partial t = \sum_{j \neq i} 
(R_{ji} \phi_j - R_{ij} \phi_i )\ ,
\label{equation:equation8}
\ee
where $R_{ij}$ is the total transition rate from state $i$ to state $j$,
including both collisional and radiative contributions.

The numerical methods used to solve the system of equations (1) through 
(8) are described in [23] and [27].  Briefly, the partial differential 
equations are converted to second order accurate finite differences in column 
depth and time.  Because of the huge range in column depth (and spatial) 
scales in the problem (length scales between the transition region and corona 
can change by 4 orders of magnitude or more!) a uniform grid is not 
possible.  Therefore a nonuniform, dynamically adjusted grid is essential in 
order that steep features remain numerically resolved.  At the end of each 
time step, the numerical resolution is checked, and zones are split and merged 
as needed.  Since the finite difference equations are centered in time, they 
must be solved implicitly.  A generalized Newton-Raphson technique is used to 
solve the difference equations at each time step.

\noindent
\underline{B.  Flare Heating, Radiative Losses, and Thermal Conduction}

The quantity $Q_{fl}$ in the energy equation represents flare heating within
the model loop.  The specific form of $Q_{fl}$ depends on the heating model
chosen.  In the case of thick target heating it is assumed that the energetic
electrons are injected near the loop apex.  For a specified input spectrum of
electrons (power law $\delta$ and low energy cutoff $E_c$) the heating rate
depends only on the column depth $N$, and has the following form [27]:
\be
Q_{tt}(N) = [(\delta -2) / 6N_c ]\  B(\delta/2,1/3) \alpha^{-\delta / 2}\ 
\{ a -b[N / (\alpha N_c )]^2 \}\ F(t)\ ,
\label{equation:equation9a}
\ee
for $N \leq \alpha N_c$, and 
\be
Q_{tt}(N) =  [(\delta-2) / 6 N_c ]\ B(\delta/2,1/3)\  [N / N_c ]^{-\delta / 2}\ 
F(t)\ ,
\label{equation:equation9b}
\ee
for $N > \alpha N_c$.  The quantity $N_c$, the stopping depth of electrons
with cutoff energy $E_c$ [KeV], is given by $N_c = E_c^2 / (3 K_{coll} )$,
where $K_{coll} = 3.64 \times 10^{-18}$ [KeV$^2$\ cm$^2$] [9].  The quantity
$B(x,y)$ is the beta function, $a=1+\delta / 4$; $b=\delta / 4$; and
$\alpha = [(\delta / 9) a B(\delta / 2, 1/3 ) ]^{2/(\delta-2)}$.  For all of
the thick target simulations discussed here, the low energy cutoff $E_c$ was
chosen to be 20 [KeV], and the power law index $\delta$ to be $4$.

In the case of thermal model heating, the heating function is assumed to be
gaussian in column depth about the loop apex:
\be
Q_{th} (N) = 2 / (\pi ^{1/2} \sigma ) e^{-(N / \sigma)^2}\ F(t)\ ,
\label{equation:equation10}
\ee
where the heating region $N \lsim \sigma$ is in the preflare corona.  In
both cases, $F(t)$ is the total energy input rate [erg\ cm$^{-2}$\ s$^{-1}$] 
into the loop as a function of time.

The radiative term $R$ in the energy equation is broken into a number of
components.  The contribution from hydrogen is computed from the solutions
to the atomic rate and radiation transport equations (\ref{equation:equation8})
and (\ref{equation:equation6}); it is given by
\be
R_H = \sum_{u > \ell} h \nu_{u \ell} (R_{u \ell} \phi_{u} - 
R_{\ell u} \phi_{\ell} )\ ,
\label{equation:equation11}
\ee
where $u,\ell = 1,2,c$.  $R_{ij}$ is the sum of both the collisional and 
radiative rates.
More details are given in [23] and [27].  It is well known that CaII and MgII 
are important contributors to radiative losses in the chromosphere, but are 
also optically thick.  Because of the large number of levels needed to 
properly describe these ions, it is not presently possible to include them in
the radiation transport portion of the code.  However, an approximate method 
of computing losses from these important ions was developed which was 
computationally expedient, yet accounted for the most important optical depth 
effects.  This is described in the appendix of [27].  The radiative losses 
from the H- continuum are adopted from HENOUX and NAKAGAWA [32], who use an 
optically thin approximation.  The radiative loss rate from the hydrogen free-
free continuum is from RICCHIAZZI [33], and is also optically thin, but 
includes an incident photospheric radiation field.  The contribution of all 
the other elemental species is assumed to be optically thin.  This loss rate 
is taken from [34], but with the hydrogen, calcium, and magnesium 
contributions removed [35], since these are calculated explicitly in the code.

The term ``$C$'' in the energy equation is the contribution to heating or 
cooling by thermal conduction, and is given by $C = -dF_c/dN$, where $F_c$ 
is the 
conductive flux.  Typically, the conductive flux is assumed to be given by [36]
\be
F_{cl} = - K(T,n_e) dT / dz\ .
\label{equation:equation12}
\ee
However, this expression for the conductive flux breaks down if the 
temperature gradient becomes too steep: the flux cannot exceed that given by 
all the local electrons moving in one direction at their thermal speed.  This 
defines an upper bound to the conductive flux
\be
| F_{sat} | = a n_e T^{3/2}\ ,
\label{equation:equation13}
\ee
where the quantity $a$ is related to the experimentally derived 
``flux limiter'' $f$ from laser ablation experiments as 
\be
a = (k_b^3 / m_e )^{1/2} f\ .
\label{equation:equation14}
\ee
Compatibility with laser ablation studies suggests that $f \sim 0.1$ [37].
As an {\it Ansatz} then, the conductive flux $F_c$ is taken to be
\be
F_c = F_{cl} / (1 + F_{cl} / F_{sat} )\ ,
\label{equation:equation15}
\ee
from which $C = - d F_c / dN$ is computed [27].
A shortcoming of these calculations 
is that heat conduction becomes an intrinsically non-local problem when the 
electron mean free path exceeds the temperature scale height (as can easily 
happen during a flare), and the contribution to local conductive heating 
depends globally on the temperature distribution.  No attempt to include such 
effects has been included in this code.  Recent progress has been made, 
however [38,39], which may allow the inclusion of nonlocal conduction effects 
into radiation hydrodynamic codes in the future.

\noindent
\underline{C.  Overview of the Numerical Simulations}

In this subsection, the overall results of the numerical simulations are 
discussed in order to impart some feeling for the most important physical 
mechanisms during the impulsive phase.  Turning first to the thick target 
model, the simulation shown in Fig. (3), one notices a number of important 
phenomena.  First of all, shortly after the onset of electron heating at 
t=0 [s], the topmost portion of the chromosphere heats up very rapidly (on a 
time scale of about 1 second) to coronal temperatures.  This seems to happen 
more or less at constant density, resulting in a tremendous overpressure in 
this region.  This overpressure quickly begins to drive upward motion of the 
heated material at speeds in excess of several hundred [km s$^{-1}$].  
At the same 
time, the overpressure also begins to drive downflows into into the remaining 
portion of the chromosphere, but at speeds generally less than 
100 [km s$^{-1}$].  
Interestingly, this downflowing region is both much denser and much cooler 
than the chromospheric material ahead of it.  The effects of heating by 
nonthermal electrons below this downward moving ``chromospheric condensation'' 
seem mainly to be the heating of the chromosphere sufficiently to ionize it, 
and to maintain a quasi-steady balance between heating by the nonthermal 
electrons and the dominant radiative loss mechanism, which in this case is 
optically thin metal losses.

\begin{figure}
\includegraphics[width=6.5in]{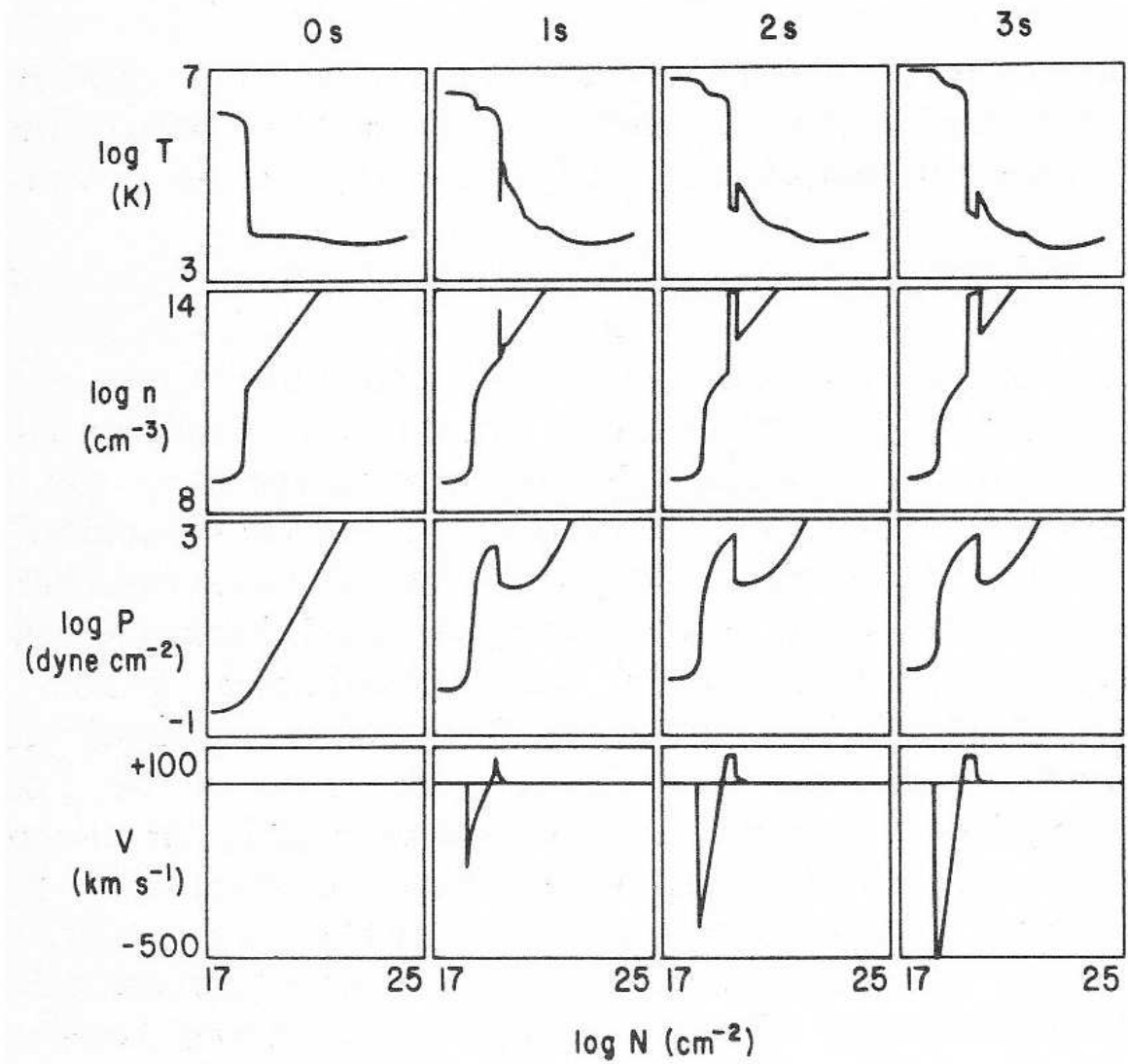}
\caption{
Evolution of 
temperature, number density, 
pressure, and velocity in 
the loop atmosphere for a 
thick target energy flux of 
$10^{11}$  [erg cm$^{-2}$ s$^{-1}$].  
The column depth $N$ is measured 
from the loop  apex.  Note 
velocities away from the 
loop apex (i.e. downward) 
are considered positive.
}
\label{fig:figure3}
\end{figure}

\begin{figure}
\includegraphics[width=6.5in]{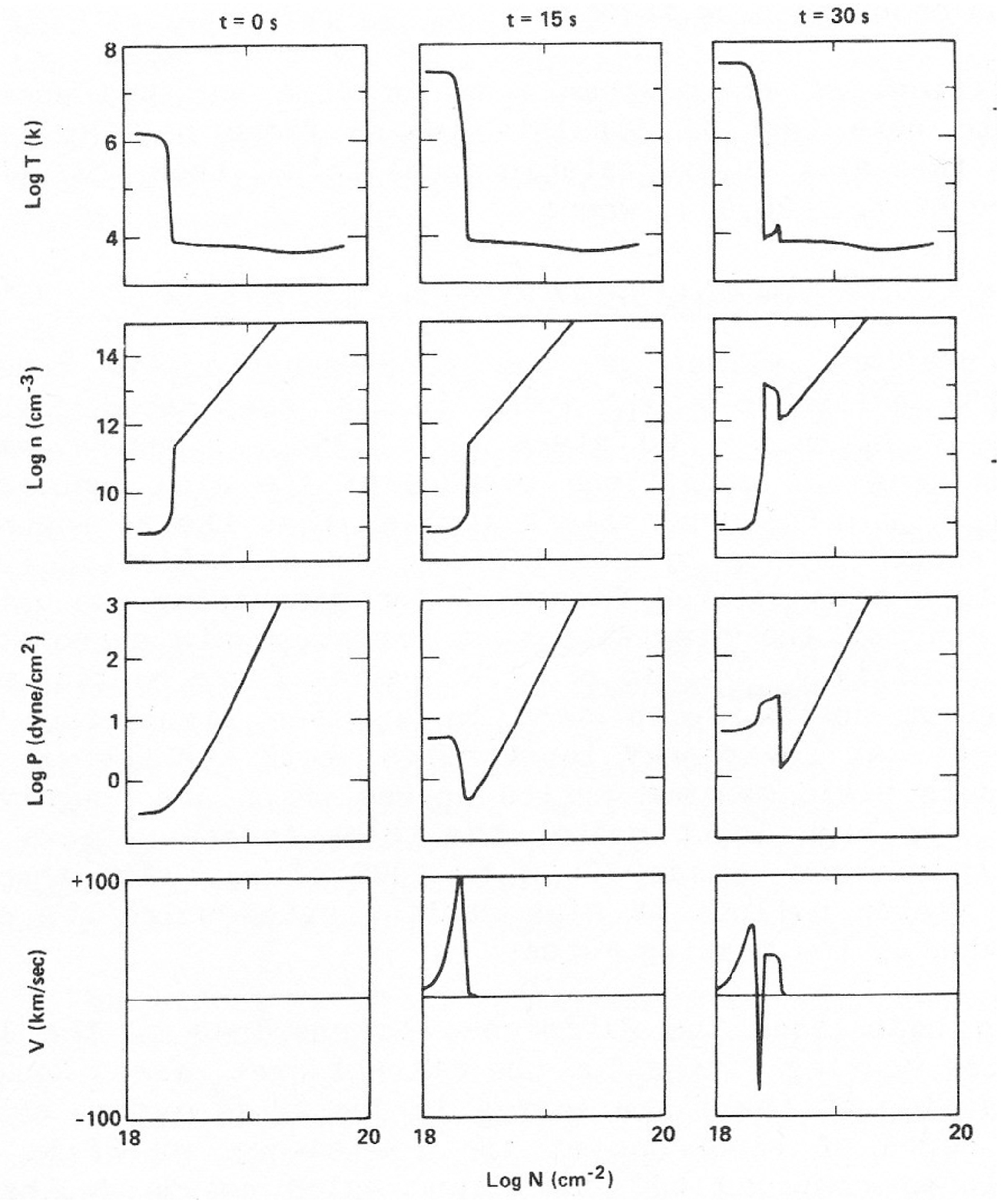}
\caption{
Evolution of 
temperature, number density, 
pressure, and velocity in 
the loop atmosphere for
thermal model heating
with an energy input flux of
$10^{9}$  [erg cm$^{-2}$ s$^{-1}$].  
The column depth $N$ is measured 
from the loop  apex.  Note 
velocities away from the 
loop apex (i.e. downward) 
are considered positive.
}
\label{fig:figure4}
\end{figure}

In the thermal model of flare heating, illustrated in Fig. (4), the 
evolution is somewhat different.  After heating commences at t=0 [s], the 
upper portion of the corona corresponding to the energy deposition region 
rapidly achieves a temperature of roughly 3x10$^7$ [K], after which the 
additional flare heating goes into driving a saturated conduction front into 
the remaining portion of the ambient corona.  It takes nearly 25 [s] for the 
conduction front to traverse the 30000 [km] coronal portion of the loop before 
finally reaching the chromosphere.  During this time, the coronal plasma is 
actually moving downward.  When the front hits the top of the chromosphere, it 
begins to rapidly heat the dense material to temperatures of roughly 
10$^7$ [K], 
resulting in a local overpressure, although not as great as that for thick 
target explosive evaporation.  This overpressure then drives upward motion of 
the heated material, reversing the earlier downward motion of coronal 
plasma.  The upflow speeds, however, are not as great as in the thick target 
case, being just over 100 [km s$^{-1}$].  The 
overpressure also drives downflow 
into the chromosphere, as with the thick target case, but the temperatures 
involved are much lower, since there is not any direct heating source in the 
chromosphere.  The amplitude of the downflow speed is also less, a result of 
the reduced overpressure.  The dominant radiative loss mechanisms in the 
chromosphere for thermal model heating seem to be optically thick emission 
from hydrogen and singly ionized calcium and magnesium.

\section*{V.  Chromospheric Evaporation}
\label{section:evaporation}

In the standard impulsive phase scenario, the observed rapid rise in the soft 
X-ray emission measure is attributed to chromospheric evaporation, the process 
of heating the upper chromosphere past the point of thermal instability; the 
temperature then quickly increases to roughly 10$^7$ [K].  As a result of this 
rapid heating, the evaporated plasma expands upward, producing the observed 
soft X-ray line blue shifts.  One very interesting issue which the numerical 
simulations can address is the extent to which computed upflow speeds match 
the upflows inferred from observed blue shifts.

In discussing calculations of chromospheric evaporation and the ensuing 
upflows, it is useful to note that if impulsive phase flare heating takes 
place in a loop which is initially in hydrostatic equilibrium, then the upflow 
velocity is bounded above by $v*$  [28,40], where
\be
v* = [(6/5)\ \ell n(n_{ch} / n_{co})]^{1/2} c_s\ ,
\label{equation:equation16}
\ee
$n_{ch}$ and $n_{co}$ are the {\it preflare} values of the chromospheric and
coronal densities, and $c_s$ is the adiabatic sound speed in the evaporated
plasma.  Using a ``typical'' value of $n_{ch} / n_{co} = 100$ gives
$v* \simeq 2.35 c_s$, a number valid over a surprisingly wide range of
conditions because of the slow functional variation of $v*$ with 
$n_{ch}/n_{co}$. The upper limit implies that the strength of any associated
coronal shocks driven by the upflows are also limited [28].
The upper limit $v*$ provides a normalized framework for discussing the upflow 
velocities obtained in many of the simulations of chromospheric evaporation 
done to date.  When the quantity $v_{max}/v*$ is plotted against flare 
heating rate 
(where $v_{max}$ is the maximum upflow speed seen in a given simulation), an 
interesting trend is seen:  At low energy input rates, both the thermal and 
thick target heating models yield maximum upflow speeds which are roughly 10 
to 20\% of $v*$.  However, at high input rates, the thick target upflows are 
rapid and explosive, with maximum speeds of 70 to 100\% of $v*$.  The thermal 
heating model, however, yields upflows at high heating rates which are only 
slightly greater than those at low heating rates.

It is fairly easy to understand the difference in response of the loop 
atmosphere to high and low heating fluxes for the thick target case.  Roughly 
speaking, a sizeable fraction of the flare energy is deposited over a column 
depth $N_c$, the stopping depth of electrons at the low energy cutoff at 
$E_c$.  
Since $N_c$ is generally large compared to transition region column depths in 
preflare loop models, a significant fraction of the energy is deposited 
directly in the chromosphere.  At low energy fluxes, the chromospheric 
temperature simply rises until radiative losses balance the increase in the 
heating rate.  Evaporation takes place when that portion of the flare energy 
deposited in the corona results in an increased conductive flux into the 
transition region, resulting in a compensating enthalpy flux in the opposite 
direction.  This scenario has been labeled ``gentle'' evaporation.  At high 
energy fluxes, however, the upper portion of the chromosphere is unable to 
radiate away the flare energy that is being deposited there.  It therefore 
heats up {\it en masse} 
to coronal temperatures, and then expands rapidly into the 
overlying corona.  This is labeled ``explosive'' evaporation.  A consequence of 
understanding this mechanism is that one can calculate the ``threshold'' 
heating 
flux for explosive evaporation to occur by simply equating the flare heating 
rate at the top of the chromosphere with the peak radiative loss rate there.  
A simple analytic model has been developed [28] which reproduces the temporal 
upflow behavior in several published simulations of explosive evaporation.  
Basically, this ``gasbag'' model assumes that the evaporated material heats up 
isochorically until a maximum temperature is reached; the plasma is then
assumed to expand isothermally in a homologous fashion.

Evaporation from thermal heating models is driven by thermal conduction 
of heat from the corona.  Since the flare energy is being deposited in the 
corona, and radiative losses there are not terribly effective, the corona 
responds by heating up and driving a larger conductive flux into the 
transition region, leading to evaporation much like that of the ``gentle'' case 
for thick target heating, except that a much larger fraction of the released 
energy goes into the evaporation process.

The observations of blue shifts and coronal temperatures derived from 
soft X-ray lines seem to lend support to the thermal model, or to ``gentle'' 
evaporation in the thick target model, if one uses a smaller low energy cutoff 
than is generally assumed.  Inferred upflow velocities range from 10 to 30\% of 
the upper limit $v*$ [40], compatible with those seen in numerical simulations
of thermal model heating or thick-target ``gentle'' evaporation.  
Information on 
the low energy cutoff is observationally inaccessible because of the 
difficulty in unfolding the hard X-ray spectrum from the thermal soft X-ray 
source below 20 [KeV].

\section*{VI.  Formation and Evolution of Chromospheric Condensations}
\label{section:condensations}

One of the most striking features seen in Figures 3 and 4 showing the response 
of the loop atmosphere to impulsive phase flare heating is the formation in 
the chromosphere of downward moving, overdense regions.  Furthermore, in the 
thick target case (Fig. 3) the downflowing dense plasma is also significantly 
cooler than its surroundings.  These ``chromospheric condensations'' form 
shortly after the onset of chromospheric evaporation, and are in fact driven 
by the evaporation process.  In the case of thick target evaporation (Fig. 3) 
the pressure in the evaporated region exceeds the overlying coronal pressure 
as well as the pressure in the chromosphere.  This results in driving both 
upward motion (evaporation) and the downward moving chromospheric 
condensation.  The hydrodynamic response for the thermal model (Fig. 4) is 
similar:  The downward moving conduction front hits the top of the 
chromosphere, suddenly creating a local pressure maximum at that point.  
Again, this drives both upflows (evaporation) and downflows (the chromospheric 
condensation), as the conduction front propagates into the chromosphere.

The structure of the chromospheric condensation is as follows:  If the 
condensation's velocity is supersonic relative to the material ahead of it, 
the leading edge of the condensation consists of a hydrodynamic shock, which 
is of order one proton-proton mean free path thick (or less if the shock is 
collisionless).  Behind the shock is a region of rapid radiative cooling.  In 
the thick target case, this region is very thin.  The shock and cooling region 
combined together form a ``radiating shock'', a structure well known in the 
astrophysics of supernova remnants, for example.  If the condensation front is 
moving subsonically, then there is no radiating shock, but there is still a 
front present, which is approximately one radiative cooling length thick.  For 
thick target heating, this cooling length is so short that the front remains 
quite thin.

In the thick target case, the radiating shock (if it exists) is followed 
by a region which is in quasi-steady energetic equilibrium, in which 
nonthermal electron heating balances radiative losses.  This region makes up 
nearly all the mass of the condensation, with the radiating shock forming only 
at the very front.  In fact, for the purposes of describing the jump in 
conditions on either side of the radiating shock ($i.e.$ from in front of the 
condensation to within it) the radiating shock itself can be completely 
ignored:  In the moving frame of the shock front, the conservation of $\rho v$ 
and 
$P + \rho v^2$
across the front as well as the constraint that thick target heating 
(per unit mass) must balance radiative losses on either side of the radiating 
shock, unambiguously specifies what the jump in hydrodynamic variables must be 
across the radiating shock.  (This same argument holds true whether the 
``condensation front'' consists of a radiating shock or not).  If the plasma is 
completely ionized on both sides of the front (which for thick target heating 
is true), and if the dominant radiative losses are optically thin (which they 
are for thick target heating) specified by a power law
($\Lambda(T) = a T^{\alpha}$), the jump 
conditions across the condensation front can be written in the following form:
\be
(v_1 / v_2 ) = ( n_2 / n_1 ) = ( T_1 / T_2 )^{\alpha} = p^{1 / \Gamma}\ ,
\label{equation:equation17}
\ee
\be
M_1 = v_1 / c_1 = \{ [p-1]/[\Gamma (1 - p^{-1 / \Gamma} )]\}^{1/2}\ ,
\label{equation:equation18}
\ee
\be
\dot N = n_1 v_1 = n_2 v_2 = n_1 c_1 M_1\ ,
\label{equation:equation19}
\ee
\be
v_d = v_1 - v_2 = M_1 c_1 (1 - p^{-1 / \Gamma} )\ ,
\label{equation:equation20}
\ee
where $p \equiv ( P_2 / P_1 )$, $\Gamma \equiv (\alpha -1) / \alpha$, and
$c_1 \equiv ( \Gamma P_1 / \rho_1 )^{1/2}$ is the long wavelength limit
of the sound speed in a plasma with optically thin losses balancing a
heating rate constant per unit mass [29].  Subscript 1 refers to material
ahead of the condensation front, and 2 to the material behind the front
(within the condensation itself).  The application of these simple jump
conditions to the thick target condensation shown in Fig. (3) describes its
instantaneous evolution quite well [29].

The chromospheric condensation associated with the thermal model shows 
some significant differences from the thick target condensation.  In the 
latter case, the conditions within most of the condensation are determined by 
quasi-steady energy balance, while the radiating shock forms only a thin layer 
at the front of the condensation.  In the thermal model, there is no direct 
release of flare energy into the chromosphere.  The unevaporated chromosphere 
remains much cooler overall, and hence has a much longer radiative cooling 
time.  In this case, the cooling portion of the radiating shock encompasses 
virtually the entire condensation, i.e. there is essentially no region of 
quasi-steady energetic equilibrium.  To complicate matters further, it is no 
longer true that the material on either side of the shock front is fully 
ionized.  Ahead of the shock, the ionized fraction $x$ is 10\% or less, while 
immediately behind the shock the atmosphere is fully ionized during the 
initial condensation evolution.  As the condensation weakens propagating into 
the chromosphere, the ionized fraction immediately behind the shock front 
begins to fall short of unity, and eventually the shock causes little change 
in the ionized fraction.  Even early on, when the condensation is moving down 
rapidly, the ionized fraction drops from unity just behind the shock to 
roughly 0.15 at the back end of the condensation, adjacent to the flare 
transition region.  The dominant radiative loss mechanism in the thermal model 
condensation varies according to position:  Immediately behind the shock 
front, Balmer continuum recombination and losses from optically thick CaII and 
MgII dominate, while at the back end of the condensation, Lyman $\alpha$ 
and Lyman
continuum losses dominate.  The dominant loss mechanism for the thermal model 
condensation therefore cannot easily be specified in an optically thin manner.

In spite of the differences between condensations in the thick target and 
thermal model cases, and the various complexities associated with the thermal 
model condensation, there are a number of simple conclusions which can be 
drawn from the simulations:\newline
 (1)  The downflow velocity of material {\it within} 
the condensation is independent 
of column depth although it is changing with time;\newline
 (2)  In both thick target and thermal model cases, the condensations 
continually slow down as they propagate into the gravitationally stratified 
atmosphere.  This is due to two effects.  One of these is inertial, i.e. the 
density of the material ahead of the condensation increases with depth.  The 
other effect is that the pressure jump across the condensation front, which 
actually drives the downflow, is decreasing as the pressure ahead of the
condensation increases with depth.  A fairly good qualitative description of 
several of the flare simulations can be obtained by assuming that the pressure 
just behind the condensation front remains constant as the front propagates 
downward.\newline
 (3)  In both heating models, the density within the condensation ($n_2$) 
is much greater than that ahead of it ($n_1$).  This 
will remain true until the very last 
stages of the condensation's downward motion.

These observations allow one to develop a simple analytical model of 
condensation dynamics which reproduces the numerical results quite well.  For 
example, using the preceding conclusions, it is straightforward to derive an 
`equation of motion' for the condensation
\be
{\dot N}^2 = (n_1 / m ) (P_2 - P_1)
\label{equation:equation21}
\ee
where $\dot N$ is the column number accretion rate of the condensation
front, $n_1$ and $P_1$ are the density and pressure ahead of the front,
which is assumed for the moment not to vary with time.  If the atmosphere ahead
of the condensation is assumed to be in hydrostatic equilibrium (an excellent
assumption for the thermal model, but somewhat questionable for the thick
target model), described by a constant gravitational scale height $H$, then
equation (\ref{equation:equation21}) can be integrated analytically.  As a
result, the time dependence of the condensation downflow velocity is found to
be
\be
v_d(t) = \pi ( H / \tau )\ {\rm cot} [ (\pi / 2 ) (t / \tau + \alpha_0 )]\ ,
\label{equation:equation22}
\ee
where ${\rm sin}^2(\alpha_0) = N_0 / N_{max}$, with $N_0 / N_{max} << 1$ in
general.  $N_0$ is the initial formation depth of the condensation,
$N_{max}$ is the stopping depth ($N_{max} = P_2 / m g$), and $\tau$
is the {\it condensation lifetime}
\be
\tau = \pi ( H / g )^{1/2}\ .
\label{equation:equation23}
\ee
The actual downflow behavior for a thermal model simulation is described at 
least qualitatively by (\ref{equation:equation22}), 
as can be seen in Fig. (5), where the analytical 
result is compared to that from a simulation.  The scale height H of 158 [km] 
is determined by an ionized fraction of 0.1 and a chromospheric temperature of 
6660 [K], consistent with the mid chromosphere of our initial atmosphere, the 
VERNAZZA, AVERETT, and LOESER model F [41].

The success of the above simple model in describing condensation dynamics 
allows the prediction of potentially observable quantities without knowing the 
detailed thermodynamic structure of the condensation itself.  Suppose, for 
example, that the H$\alpha$ red wing asymmetry often observed 
during flares is a 
measure of the condensation downflow speed $v_d(t)$, as has been proposed by 
ICHIMOTO and KUROKAWA [42].  Then equation (\ref{equation:equation22}) 
predicts the temporal 
evolution of the velocity determined from the red wing asymmetry, and that the 
asymmetry should die away on a time scale of $\tau$, which 
is about 1 minute for 
typical chromospheric values of the temperature and ionized fraction.  
Although time resolution of H$\alpha$ asymmetries during flares 
is not sufficiently 
high to compare with the detailed temporal behavior of the simulations or with 
the analytical model, it is sufficiently good to show that the asymmetry 
{\it decay} 
time is roughly consistent with that predicted by (\ref{equation:equation23}).  
Studies are presently 
under way [43] to develop more realistic analytical models of condensation
dynamics which relax some of the assumptions used in deriving 
(\ref{equation:equation22}), but 
preliminary results show that the resulting decay times are fairly close to 
those given by 
(\ref{equation:equation23}), 
and that the initial evolution of the condensation is 
still described fairly well by (\ref{equation:equation22}).  
Observational studies are also underway 
to study the evolution of red wing asymmetries during flares with better time 
resolution [44].

\begin{figure}
\includegraphics[width=6.5in]{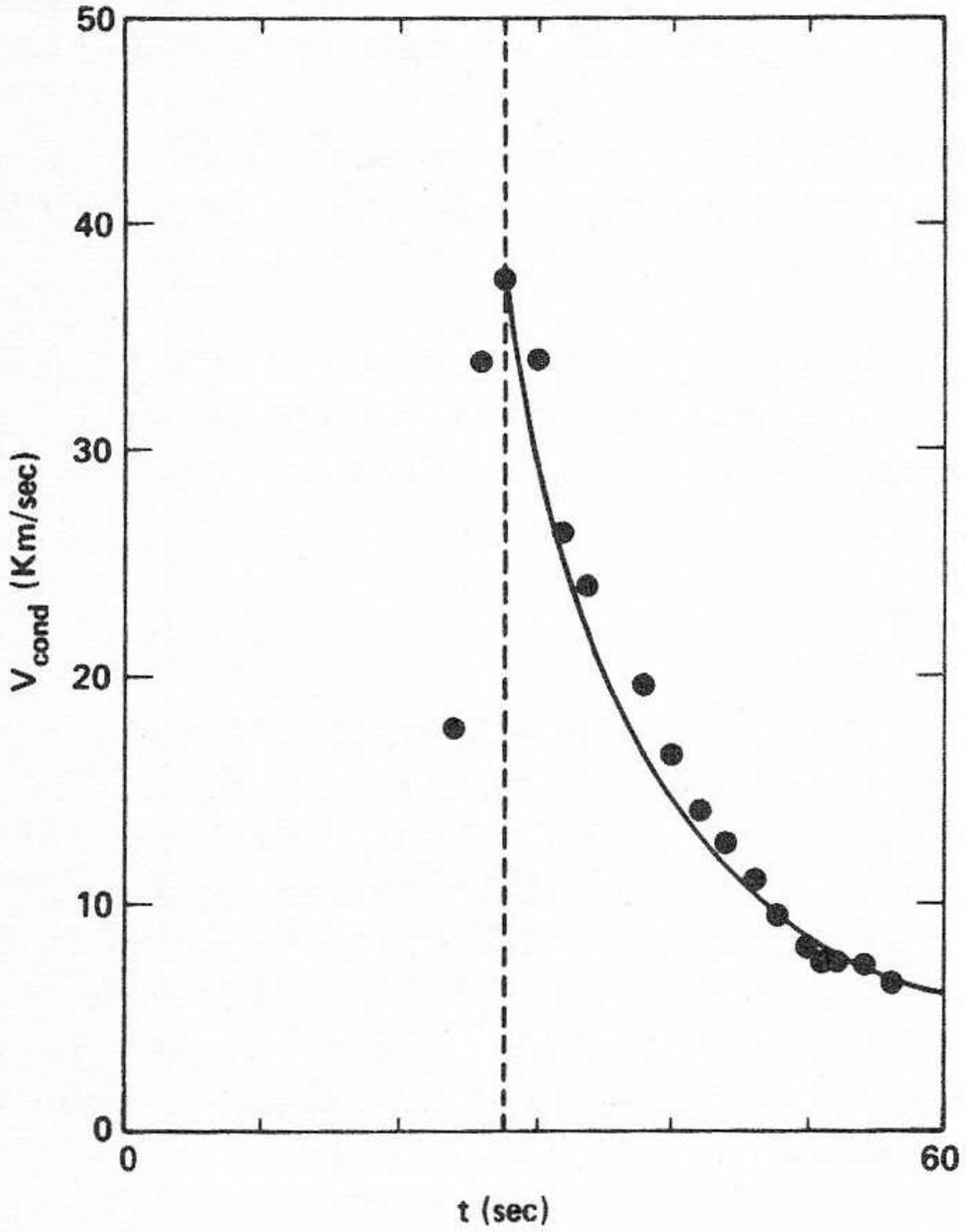}
\caption{
Comparison of 
condensation downflow speed 
computed in the thermal model 
simulation with the result 
(\ref{equation:equation22}).  The dotted line 
corresponds to t=0  [s] in 
(\ref{equation:equation22}).  The long delay before that 
corresponds to the motion of the 
conduction front through the 
corona.
}
\label{fig:figure5}
\end{figure}

\section*{VII.  Radiating Shocks}
\label{section:radshocks}

In the preceding section the precise structure of the chromospheric 
condensation was glossed over, as such detailed knowledge is not necessary in 
order to understand the overall dynamics.  However, the structure of the 
condensation, in particular the radiating shock which may form at its leading 
edge, can lead to distinct emission properties which could be used as a 
diagnostic of the temperature and density profiles in the radiating shock, in 
turn yielding information on energy deposition in the flaring atmosphere.  The 
main observational consequences of radiating shocks, if they form, is that 
they can provide a significant source of UV radiation in the impulsive phase 
of flares.  In fact, as will be illustrated shortly, a radiating shock can 
produce much more of the emission normally associated with the transition 
region than does the transition region itself, at least during the transient 
period the radiating shock exists.  To demonstrate this, the following section 
deals with the calculation of the differential emission measure $\xi(T)$ in a 
radiating shock formed by thick target heating, and compares this with 
$\xi(T)$
from the transition region.  In both cases, a simplified radiative loss rate
is assumed.

The emission flux  F from an optically thin, collisionally excited 
process in a plasma with a non-uniform distribution of temperature and 
electron density can often be expressed in the form
\be
F = \int (g(T) / T) \xi(T) dT\ ,
\label{equation:equation24}
\ee
where $g(T)$ absorbs the details of the atomic physics of the given emission 
process, and $\xi(T)$ is the differential emission measure
\be
\xi(T) = n_e^2 (d\,\ell n T / dz )^{-1}\ .
\label{equation:equation25}
\ee
Because of this relationship between a measurable quantity $F$, a supposedly 
known quantity $g(T)$, and the model dependent quantity $\xi(T)$, the 
differential emission measure is an important contact point between theory and 
observation.

The structure of the transition region is often assumed to be determined 
by a balance between conductive heating and radiative losses, and as such 
yields the following behavior for $\xi(T_5) = \xi(T/10^5[K])$:
\be
\xi_{tr} (T_5) [{\rm cm}^{-5}] = 1.16 \times 10^{27} P\ T_5^{-3/4}\ \ \ 
{\rm for\ } T_5 < 1,
\label{equation:equation26a}
\ee
\be
\xi_{tr} (T_5 ) [{\rm cm}^{-5}] = 1.16 \times 10^{27} P\ T_5^{3/2} 
[1 + 9 (T_5^{1/2} -1) ]^{-1/2} {\rm\ \  for\ } T_5 > 1,
\label{equation:equation26b}
\ee
where $P$ is the transition region pressure [dyne\ cm$^{-2}$], and the 
optically thin radiative loss function is assumed to have the form
\be
\Lambda(T_5) [{\rm erg\ cm}^3\ {\rm\ s}^{-1}] = 
7 \times 10^{-22}\ T_5^3\ {\rm for\ }
T_5 < 1,
\label{equation:equation27a}
\ee
\be
\Lambda(T_5) [{\rm erg\ cm}^3\ {\rm\ s}^{-1}] = 
7 \times 10^{-22}\ T_5^{-1}\ {\rm for\ }
T_5 > 1,
\label{equation:equation27b}
\ee
which has been demonstrated to be a reasonable approximation to the actual 
radiative losses during thick target heating [25,29].  In fact, 
(\ref{equation:equation26a}-\ref{equation:equation26b}) are an 
upper limit to $\xi_{tr}(T)$, for a given value of the pressure, 
since the flare transition region is undergoing 
evaporation, a process which must reduce $\xi_{tr}(T)$ [45].

In a radiating shock, it is straightforward to show that $\xi(T)$ is
approximately of the form
\be
\xi_{rs}(T) \simeq 5 k \dot N T / \Lambda(T),\ {\rm for\ }
T_{cond} < T < T_{max}\ ,
\label{equation:equation28}
\ee
where $T_{max}$ is determined by the strength 
of the shock and the temperature just 
ahead of it, and $T_{cond}$ by the quasi-steady 
equilibrium condition in the bulk 
of the condensation.  The quantity $\dot N$ is the instantaneous 
column number flux 
through the radiating shock.  By applying values of $\dot N$ obtained from the 
simulations, one finds that $\xi_{rs}(T) >> \xi_{tr}(T)$ at a given 
instant in time for 
the allowed temperature range within the radiating shock.  Therefore, emission 
from the radiating shock will dominate that from the transition region, and 
should therefore be observable.  There are a number of further predictions one
can make.  As the condensation decelerates into the gravitationally stratified 
chromosphere, the shock strength will weaken, reducing the peak temperature 
achieved just behind the hydrodynamic shock.  The emission from the highest 
temperatures in the radiating shock will therefore `wink out' at some early 
point in time, with emission from successively lower temperatures disappearing 
at progressively later times.  This behavior, if it in fact exists, should be 
clearly visible by an instrument such as the (recently crippled) Ultraviolet 
Spectrometer Polarimeter (UVSP) on board the Solar Maximum Mission.

In order to demonstrate this effect, and to show how the evolution of the 
radiating shock and condensation are tied together, one can combine the 
analytic behavior for the condensation downflow discussed in the previous 
section (to get an expression for $\dot N$) with the 
expression (\ref{equation:equation28}) for $\xi_{rs}(T)$.  
The condition that the atmosphere ahead of the radiating shock is in 
quasi-steady 
equilibrium between flare heating and radiative losses unambiguously 
determines the temperature just ahead of the radiating shock.  Knowing also 
the condensation downflow speed unambiguously determines the Mach number, and 
hence the maximum temperature reached in the radiating shock at each point in 
time.  It is therefore possible to predict the entire evolution of 
$\xi_{rs}(T)$ over 
the life of the radiating shock.  (In the case illustrated in Fig. (6), it is 
21 seconds.  The subsequent condensation motion (lasting $\sim\ 40$ [s]) is 
subsonic).  Three curves of $\xi(T)$ have been 
plotted in Fig. (6), corresponding 
to a specific case of thick target explosive evaporation.  The dotted curve 
($\xi_{rs}(T)$) is the differential emission measure from 
the radiating shock when it 
first forms.  The dashed curve ($\xi_{tr}(T)$) is from the transition 
region at the 
same pressure as the condensation, and the solid curve ($\bar{\xi}_{rs}(T)$) 
is the average 
from the radiating shock over the length of time the radiating shock exists.  
One concludes that:\newline
 (1) At high time resolution, the radiating shock emission should be clearly 
visible over that from the transition region at all temperatures below the 
shock maximum, but\newline
 (2) At low time resolution $\xi_{tr}(T)$ dominates over 
$\bar{\xi}_{rs}(T)$ for temperatures 
above $10^5$ [K].  Nevertheless, at low temperatures (below 
$10^5$ [K]) $\bar{\xi}_{rs}(T)$ still 
dominates that from the transition region, and causes a much steeper 
temperature dependence of $\xi(T)$: $\bar{\xi}_{rs}(T) \propto T^{-3.0}$.  
One concludes that UV 
instruments of roughly 10 second time resolution should be able to detect 
emission from the lower temperature portion of a radiating shock, and 
instruments with time resolution of 1 second or better should be able to 
detect the short lived emission from higher temperatures in radiating shocks.

\begin{figure}
\includegraphics[width=6.5in]{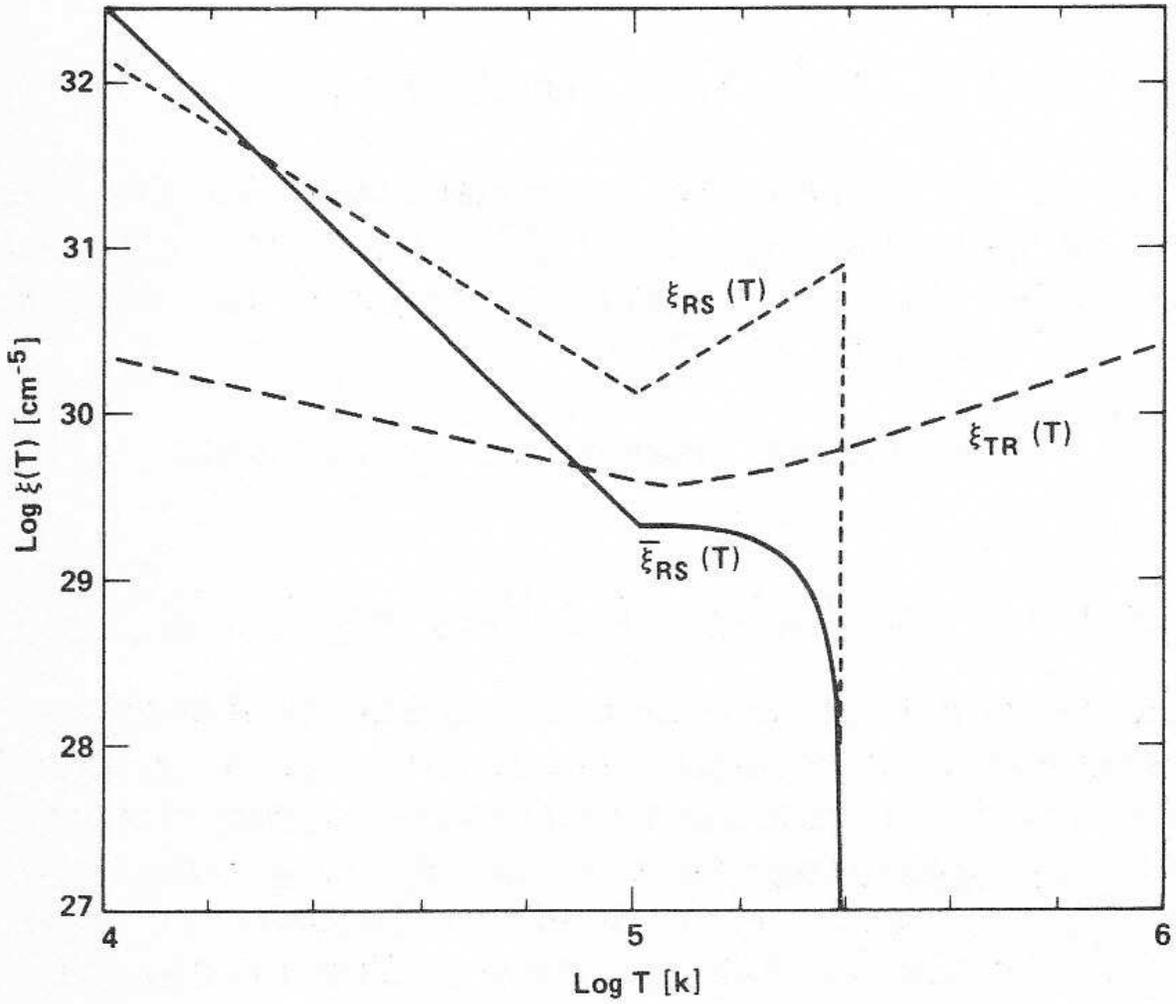}
\caption{
The dashed, 
dotted, and solid curves 
correspond to the 
differential emission 
measure from the transition 
region, the initial state of 
the radiating shock, and the 
average over the life of the 
radiating shock, 
respectively, for thick 
target heating similar to 
that used in Fig. 3.
}
\label{fig:figure6}
\end{figure}

One significant piece of physics ignored in the entire preceding 
discussion is the question of ionization equilibrium.  As is well known [46], 
the cooling time for an individual ion in a radiating shock can be shorter 
than its ionization equilibrium time scale, requiring a fully time dependent 
ionic species calculation to self-consistently determine the cooling rate in a 
radiating shock.  Many such calculations have been done for radiating shocks 
in supernova remnants, for example [46], but at densities 11 or 12 orders of 
magnitude lower than the solar flare case.  In the SNR case, important 
contributors to the loss rate include many forbidden and semi-forbidden 
transitions which would probably be quenched in the solar case.  It is clear 
that calculations of the type performed by SHULL and MCKEE [46] need to be 
done at much higher densities, and need to include the rapid weakening of the 
shock with time in order to make realistic predictions of specific UV spectral 
features.  However, a study of non-equilibrium ionization effects in flares 
undertaken by MACNEICE et al [19] showed that the actual temperature structure 
in such a self-consistent calculation should not differ too much from that 
reflected by the $\xi(T)$ calculations presented here using an ionization 
equilibrium radiative loss rate.

Radiating shocks in thermal model calculations seem to be significantly 
different from thick target radiating shocks.  The chromospheric pressure 
enhancement due to evaporation by the conduction front in general seems to be 
much less than that for explosive evaporation. [16,30].  This results in a 
slower downflow speed, and hence a lower postshock temperature.  In addition, 
the material ahead of the condensation is generally not ionized, as it has not 
been preheated by the nonthermal electrons.  A great deal of the shock energy 
goes into simply ionizing the hydrogen, rather than heating the shocked 
material to high temperatures.  As mentioned earlier, the characteristic 
emission in the thermal model radiating shock tends to be Balmer continuum 
recombination and optically thick CaII and MgII line radiation close to the 
shock front itself, and Lyman $\alpha$ and 
Lyman continuum radiation further toward 
the back (top) of the radiating shock.  There is no natural distinction 
between the radiating shock and the rest of the chromospheric condensation, as 
there is in the thick target case.

\section*{VIII.  Conclusions}
\label{section:conclusions}
The motivation for performing detailed calculations of the radiative 
hydrodynamic response of loop models to flare energy release is that flare 
observations suggest that such models might have some realistic basis.  As a 
result of performing these calculations, however, the situation has been 
turned around, and it is now possible to make new predictions of potentially 
observable phenomena based on these calculations.  For example, if the thick 
target model of loop heating is to be taken at face value, then one expects
that:\newline
(1) Explosive evaporation at large electron energy fluxes will initiate a 
downward moving chromospheric condensation, producing observable red 
asymmetries in H$\alpha$, and the subsequent decay of these asymmetries on 
time scales of $\tau =  \pi (H/g)^{1/2}$;\newline
 (2) Simultaneous with the initial downflow of a condensation, a bright 
radiating shock will form, emitting transition region-like UV radiation, but 
with the higher temperature emission rapidly `winking out' as the shock 
decelerates.  If it is possible to measure the downflow in the condensation 
sufficiently accurately, one should furthermore be able to link given downflow 
speeds with the disappearance of specific UV transitions as the condensation 
slows down.\newline
 (3) The temperature dependence of the time averaged differential emission 
measure below $10^5$ [K] should be overall somewhat steeper during flares than 
the quiet Sun, where there presumably are no radiating shocks present.

\acknowledgements
This work was performed at the Lawrence Livermore National Laboratory under
the auspices of the US Department of Energy under contract No. W-7405-ENG-48.

\noindent
\underline{Discussion:}

\noindent
{\bf Frank H. Shu:}  You talked mostly about the condensation phase behind the 
radiating shock, but this must be followed by an evaporative phase in which 
thermal conduction eats away at the condensation.  How long does this take, 
and is it observable?

\noindent
{\bf Fisher:}  
It is certainly true that the back end (top) of the condensation is 
being evaporated conductively.  However, the column number accretion rate of 
the condensation (roughly $10^{19}$ [particles cm$^{-2}$s$^{-1}$] ) 
is much greater than the 
conductive evaporation rate 
(roughly $10^{17}$ [particles cm$^{-2}$s$^{-1}$]).  Basically, 
the condensation has already stopped by the time a significant fraction of it 
has been evaporated.  The evaporation definitely has an observable 
consequence, namely the increase of the soft X-ray emission measure.  Observed 
time scales for its increase are roughly consistent with those obtained with 
estimates of the coronal conductive flux.

\noindent
{\bf Rino Bandiera:}  When the chromosphere is heated up, could the gas energy 
density be higher than the magnetic one?  If yes, could the magnetic field 
structure be affected, and in which way?

\noindent
{\bf Fisher:}  
The largest gas pressure encountered in any of the simulations of 
explosive evaporation was about 400 [dyne cm$^{-2}$], corresponding to an 
equivalent magnetic field of about 100 [gauss].  This is not an unthinkable 
value, as active region loops have been observed with similar estimated field 
strengths.  In such a case, I imagine that the evaporated region might 
initially bulge out, resulting in Alfven waves driven along the loop.  In 
addition, magnetoacoustic waves might be driven out in a perpendicular 
direction into the adjacent flux tubes.  The dissipation of the energy in 
these ways would reduce that available for mass flow along the loop, and the 
upflow speeds would probably be reduced.  However, there is some evidence that 
compact flare loops have much higher field strengths than this, in which case 
the rigid tube approach is still a good approximation.

\noindent
{\bf Vincent Icke:}  I'm worried about your geometry.  In double ribbon flares 
there's lots of evidence that the magnetic configuration changes rapidly, so 
that new material is being bombarded all the time.  Does that not require 
changes in your approach?

\noindent
{\bf Fisher:}  I presume 
you're referring to the expansion of the H$\alpha$ ribbons, and 
the interpretation of that in the Kopp and Pneumann reconnection picture, 
where new loops are continuously forming through the reconnection of open 
field lines.  In that case, I would expect that the simple closed flux tube 
model I discussed here would work once a closed flux tube had been formed, but 
would be irrelevant before that.  One would then need to somehow superimpose a 
collection of these 1-d loop models in order to describe the evolving arcade.

\noindent
{\bf Vincent Icke:}  Would 
you agree that the radio observations favor the particle 
injection (thick target) model over the thermal model?

\noindent
{\bf Fisher:}  Yes.  During the impulsive phase, the observation of what is 
apparently gyrosynchrotron emission in the microwave spectrum, and type III 
radio bursts, apparently due to beams of electrons accelerated into the 
interplanetary medium, argue strongly in favor of particle acceleration.

\noindent
{\bf Reuven Opher:}  The damping rate of 
turbulent Alfven waves is proportional to 
$\nu k^2/n$, where $\nu$ is the collision frequency, 
$n$ is the number density, and $k$ is 
the average wave number determined by the turbulent spectrum which can be 
fairly large.  Appreciable absorption can occur in the corona and in the 
transition region.  How do these waves affect your model and what are the 
observable consequences?

\noindent
{\bf Fisher:}  The absorption of Alfven waves has been proposed as a heating 
mechanism in flares by Emslie and Sturrock.  They were most interested in 
exploring the possibility that the observed heating of the temperature minimum 
region might be explained by Alfven waves.  Emslie now says he believes that 
although this is possible, one must pick a contrived set of parameters for it 
to work properly.

\noindent
{\bf Reuven Opher:}  The proposed mechanism for producing the fast electrons is 
magnetic reconnection which is occuring in a highly turbulent region.  This 
region should also produce a large fraction, or a major fraction, of its 
energy in turbulent Alfven waves which will propagate at the Alfven 
velocity.  For a typical loop dimension they should reach the transition 
region on the order of a second, which is in agreement with the Oxygen V line 
data.

\noindent
{\bf Miguel H. Ibanez S.:}  Have you checked whether 
thermal and magneto-acoustic 
waves play some role in your models?

\noindent
{\bf Fisher:}  A self-consistent 
treatment of magneto-acoustic waves in these 1-d 
models has essentially been precluded by assuming that the loop is a rigid 
tube.  One can include the effects in an ad-hoc manner by specifying (for 
example) what heating effects one expects the waves to have, but this has not 
been done in these models.  As far as thermal waves in 1-d loop models goes, a 
considerable amount of effort has gone into the study of the stability of 
thermal waves in solar loops.  You might want to look at the work by Antiochos 
and by McClymont and Craig.

\noindent
{\bf {\AA}ke Nordlund:}  I don't want to 
be negative over a good talk, and good work, 
but the statement that ``The probabilistic radiative transfer has been tested 
and has been found to do quite a good job'' would need some additional
comments.  I am skeptical about P.R.T. in situations with super Doppler 
velocity fields.

\noindent
{\bf Fisher:}  You're quite right in being skeptical in the case of velocity 
gradients in optically thick material.  We employ escape probabilities that 
are computed using the static atmosphere assumption.  The validity of this is 
tested in the numerical code using criteria formulated by Hummer and Rybicki, 
and we find that in nearly all cases $dv/d\tau$ in 
hydrogen transitions is small 
enough that the static assumption is OK.  However, I expect that the escape 
probabilities in use for CaII and MgII transitions are probably too low, for 
this reason.  That means the CaII and MgII losses are probably underestimated 
in the region just behind the radiating shock, for example.

\noindent
{\bf Dmitri Mihalas:}  How important is 
escape of radiation out the sides of your 
loop model as compared to escape out the top of the chromosphere?

\noindent
{\bf Fisher:}  Well, that depends on 
the ratio of the width of the chromospheric 
portion of a compact flare loop to its height, which I don't think is terribly 
well known.  My guess, from the size of bright H$\alpha$ kernels during 
the impulsive 
phase, is that the width of such loops is roughly the same as the 
chromospheric depth, namely 1000 or 2000 km.  It probably varies from loop to 
loop.  This effect is not included in the numerical models.

\noindent
{\bf John Brown:}  Does the creation of condensations, 
and the existence of a 
threshold for explosive evaporation, depend on the instantaneous switch-on of 
your beam?  Would they vanish if the beam were ramped up gradually?

\noindent
{\bf Fisher:}  Given 
an average heating rate (per particle) $Q_0$, a column thickness 
of the chromosphere $N_0$ unable to radiate away the flare heating, with the 
spatial extent of this region being $L_0$, then the timescale necessary to 
achieve explosive evaporation ($\tau_{exp}$) is 
$\tau_{exp} \sim  [m L_0^2/(2Q_0)]^{1/3}$, where m is 
the mean mass per hydrogen nucleus.  If $\tau_{exp} > \tau_{rise}$, 
where $\tau_{rise}$ is the rise 
time for the beam energy flux, then I expect explosive evaporation to occur, 
and to initiate the formation of a condensation.  It should be possible, with 
spatially and temporally resolved hard X-ray observations, to estimate both of 
these time scales.  If one can indeed detect condensations through H$\alpha$ 
asymmetries, then it should be possible to make a consistency check of this 
whole scenario by seeing if the existence of red wing asymmetries depends on 
the above time scale inequality.

\noindent
{\bf John Brown:}  Regarding the numerical techniques, could you summarize the 
current state of the SMM hydrodynamic `benchmark' exercise in which there were 
serious discrepencies between different codes run for the same problem?

\noindent
{\bf Fisher:}  There were indeed serious discrepencies between the different 
codes.  It turned out some of these were due to what seemed initially to be 
minor details such as using different assumptions about the ambient heating 
function used to keep the initial atmosphere in energetic equilibrium.  Other 
discrepencies might be explained by insufficient numerical resolution of the 
transition region in some of the models.

\noindent
{\bf John Brown:}  In response to the question by Rino Bandiera:  It has been 
suggested by A. G. Emslie that energy deposition by the beam could indeed 
produce a gas pressure exceeding the magnetic pressure, and that the resulting
flux tube expansion could act as a trigger on neighboring tubes and initiate 
subsequent further reconnection and acceleration.

In response to the question by Reuven Opher on Alfven waves, Emslie and 
Sturrock studied Alfven wave heating.  While it is possible, for carefully 
chosen parameters, to heat the chromosphere in this way, the process is too 
slow (for the observations) at layers much below the transition region because 
the Alfven speed becomes very low there.
     Two final comments I would make, of possibly more general interest here 
are:\newline
 (1)  This paper is the first mention we have had of nonthermal particles as 
an important element in the hydrodynamic equations.  They may also be relevant 
in accretion columns and around QSO's.\newline
 (2)  If the flux tube is not uniform in cross-section, there arises the 
important effect of loss-cone instability and generation of radiation by 
masering (Melrose and Dulk).  Such radiation (decimetric for solar parameters) 
plays the role of transporting energy to (and hence driving motions in) the 
plasma outside the tube, hence turning a 1-d problem into a 3-d problem.  This 
process may also have analogues in cosmic ray problems.


\begin{thebibliography}{50}

\bibitem{Krieger}
1.   Krieger, A.S.: "X-ray Observations of Solar Structural Features", in 
     Proceedings of the OSO-8 Workshop, (Laboratory for Atmospheric and 
     Space Sciences, University of Colorado 1977)

\bibitem{Svestka}
2.   Svestka, Z.:  Solar Flares, (D. Reidel Publ. Co., Dordrecht, Holland
     (1976)

\bibitem{Pneumann}
3.  Pneumann, G.W.: "Two Ribbon Flares: (post) Flare Loops", in Solar Flare 
    Magnetohydrodynamics, E.R. Priest (ed), (Gordon and Breach Publ., New 
    York 1981)

\bibitem{Kopp}
4.  Kopp, R.A., Pneumann, G.W.: Solar Physics 50, 85. (1976)

\bibitem{Woodgate}
5.  Woodgate, B.E., Shine, R.A., Poland, A.I., Orwig, L.E.: Astrophysical 
    Journal 265, 530. (1983)

\bibitem{CG}
6.  Canfield, R.C., Gunkler, T.A.: Astrophysical Journal 288, 353. (1985)

\bibitem{Kane}
7.  Kane, S.R., Donnelly, R.F.: Astrophysical Journal 164, 151. (1971)

\bibitem{AGD}
8.  Antonucci, E.,Gabriel, A.H., Dennis, B.R.: Astrophysical Journal 287, 
     917. (1984)

\bibitem{Brown}
9.  Brown, J.C.: Solar Physics 26, 441. (1972)

\bibitem{LH}
10.  Lin, R.P., Hudson, H.S.: Solar Physics 50, 153. (1976)

\bibitem{KP}
11.  Kostyuk, N.D., Pikel'ner, S.B.: Soviet Astronomy, 18, 590. (1975)

\bibitem{Kostyuk}
12.  Kostyuk, N.D.: Soviet Astronomy, 20, 206. (1976)

\bibitem{CM}
13.  Craig, I.J.D., McClymont, A.N.: Solar Physics 50, 133. (1976)

\bibitem{LBKK}
14.  Livshitz, M.A., Badalyan, O.G., Kosovichev, A.G., Katsova, M.M.: Solar 
     Physics 73, 269. (1981)

\bibitem{SSS}
15.  Somov, B.V., Syrovatskii, S.I., Spektor, A.R.: Solar Physics 73, 145. 
     (1981)

\bibitem{Cheng}
16.  Cheng, C.C., Oran, E.S., Doschek, G.A., Boris, J.P., Mariska, J.T.: 
     Astrophysical Journal 265, 1090. (1983)

\bibitem{Palla}
17.  Pallavicini, R., Peres, G., Serio, S., Vaiana, G., Acton, L., 
     Leibacher, J., Rosner, R.:  Astrophysical Journal 270, 270. (1983)

\bibitem{Duij}
18.  Duijveman, A., Somov, B.V., Spektor, A.R.: Solar Physics 88, 257. (1983)

\bibitem{MacNeice}
19.  MacNeice, P., McWhirter, R.W.P., Spicer, D.S., Burgess, A.: Solar 
     Physics 90, 357. (1984)

\bibitem{CKD}
20.  Cheng, C.C., Karpen, J.T., Doschek, G.A.: Astrophysical Journal 286,
     787. (1984)

\bibitem{MAVN}
21.  Machado, M.E., Averett, E.H., Vernazza, J.E., Noyes, R.W.: Astrophysical 
     Journal 242, 336. (1981)

\bibitem{RC}
22.  Ricchiazzi, P.J., Canfield, R.C.: Astrophysical Journal 272, 739. (1983)

\bibitem{PaperI}
23.  McClymont, A.N., Canfield, R.C.: Astrophysical Journal 265, 483 (1983)

\bibitem{CFM}
24.  Canfield, R.C., Fisher, G.H., McClymont, A.N.: Astrophysical Journal 
     289, 507. (1983)

\bibitem{SandyDick}
25.  McClymont, A.N., Canfield, R.C.: Astrophysical Journal 265, 497. (1983)

\bibitem{An}
26.  An, C.H., Canfield, R.C., Fisher, G.H., McClymont, A.N.: Astrophysical 
     Journal 267, 421. (1983)

\bibitem{FCMa}
27.  Fisher, G.H., Canfield, R.C., McClymont, A.N.: Astrophysical Journal 
     289, 414. (1985)

\bibitem{FCMb}
28.  Fisher, G.H., Canfield, R.C., McClymont, A.N.: Astrophysical Journal 
     289, 425. (1985)

\bibitem{FCMc}
29.  Fisher, G.H., Canfield, R.C., McClymont, A.N.: Astrophysical Journal 
     289, 434. (1985)

\bibitem{noop}
30.  Fisher, G.H.: In prep. (1985)

\bibitem{Prob}
31.  Canfield, R.C., McClymont, A.N., Puetter, R.C.:  "Probabilistic 
     Radiative Transfer", in Methods in Radiative Transfer, ed. Wolfgang 
     Kalkofen (Cambridge Univ. Press, 1984)

\bibitem{HN}
32.  Henoux, J.C., Nakagawa, Y.: Astronomy and Astrophysics 57, 105. (1977)

\bibitem{PJRthesis}
33.  Ricchiazzi, P.J.: PhD Thesis, University of California at San Diego 
     (1982)

\bibitem{RCS}
34.  Raymond, J.C., Cox, D.P., Smith, B.W.: Astrophysical Journal 204. 290. 
     (1976)

\bibitem{Rayprivate}
35.  Raymond, J.C.: Private communication to P. J. Ricchiazzi (1980)

\bibitem{Spitzer}
36.  Spitzer, L.: Physics of Fully Ionized Gases (Interscience, New York 1962)

\bibitem{Hauer}
37.  Hauer, A., Mead, W.C., Willi, O., Kilkenny, J.D., Bradley, D.K., 
     Tabatabaei, S.D., Hooker, C.: Physical Review Letters, 53, 2563. (1984)

\bibitem{Luciani}
38.  Luciani, J.F., Mora, P., Pallat, R.: Physics of Fluids, 28, 835. (1985)

\bibitem{KD}
39.  Karpen, J.T., Devore, C.R.: Astrophysical Journal, submitted. (1985)

\bibitem{FCM84}
40.  Fisher, G.H., Canfield, R.C., McClymont, A.N.: Astrophysical Journal 
     (Letters) 281, L79 (1984)

\bibitem{VAL}
41.  Vernazza, J.E., Averett, E.H., Loeser, R.: Astrophysical Journal 
     (Supplement) 45, 619. (1981)

\bibitem{IK}
42.  Ichimoto, K., Kurokawa, H.: Solar Physics 93, 105. (1984)

\bibitem{noop2}
43.  Fisher, G.H.: In prep. (1985)

\bibitem{Dickprivate}
44.  Canfield, R.C.: Private communication. (1985)

\bibitem{IanSandy}
45.  Craig, I.J.D., McClymont, A.N.: Astrophysical Journal (submitted) (1985)

\bibitem{SM79}
46.  Shull, J.M., McKee, C.F.: Astrophysical Journal 227, 131 (1979)

\end{thebibliography}

\end{document}